\documentclass[aps,showpacs,preprintnumbers,amsmath,amssymb]{revtex4}
\usepackage{dcolumn}
\usepackage[dvips]{epsfig}

\usepackage{float}
\usepackage{graphics,epsfig}
\usepackage{graphicx}
\usepackage{dcolumn}
\usepackage{bm}
\usepackage{gensymb}
\usepackage{subfigure}

\begin{document}
\baselineskip=0.8 cm
\title{\bf Kerr Black hole shadows in Melvin magnetic field with stable photon orbits}

\author{Mingzhi Wang$^{1}$\footnote{Corresponding author: wmz9085@126.com},  Songbai Chen$^{2,3}$\footnote{csb3752@hunnu.edu.cn},  Jiliang
Jing$^{2,3}$\footnote{jljing@hunnu.edu.cn}
}
\affiliation{$ ^1$School of Mathematics and Physics, Qingdao University of Science and Technology, Qingdao, Shandong 266061, People's Republic of
China \\ $ ^2$Institute of Physics and Department of Physics, Key Laboratory of Low Dimensional Quantum Structures
and Quantum Control of Ministry of Education, Synergetic Innovation Center for Quantum Effects and Applications,
Hunan Normal University, Changsha, Hunan 410081, People's Republic of China\\
$ ^3$Center for Gravitation and Cosmology, College of Physical Science and Technology,
Yangzhou University, Yangzhou 225009, China}

\begin{abstract}
\baselineskip=0.6 cm
\begin{center}
{\bf Abstract}
\end{center}
We have studied the spacetime of a Kerr black hole immersed in Melvin magnetic field, and found not only unstable light rings could exist, but also stable light rings could exist. Both the prograde and retrograde unstable light rings radiuses increase with the magnetic field parameter $B$, but it is the opposite for stable light rings. The existence of unstable, stable light rings depend on both the rotation parameter $a$ and the magnetic field parameter $B$. For a certain $a$, there are both the prograde and retroprade unstable (stable) light rings when $B$ is less than a critical value $B_{c}$ of retrograde light ring. In this case, the shadows of Melvin-Kerr black hole have two gray regions on both sides of the middle main shadow, which correspond to the prograde and retrograde stable photon orbits. The photons in stable orbits are always moving around Melvin-Kerr black hole, they can't enter the black hole or escape to infinity. As $B$ continues to increase, there is only the prograde unstable (stable) light ring. In this case, the gray region only emerges in the life of the main shadow, which corresponds to the prograde stable photon orbits. The absence of the retrograde unstable (stable) light rings makes the Melvin-Kerr black hole shadow an half-panoramic (equatorial) shadow. When $B$ is bigger than $B_{C}$ of prograde light ring, neither prograde nor retroprade unstable (stable) light rings exist. In this case, the shadow of Melvin-Kerr black hole has no gray region for stable photon orbits, and becomes a panoramic (equatorial) shadow. In addition, there also exist some self-similar fractal structures in the shadow of Melvin-Kerr black hole arising from the chaotic motion of photon.

\end{abstract}

 \maketitle
\newpage
\section{Introduction}

The first image of the supermassive black hole in the center of the giant elliptical galaxy M87 has been announced by Event Horizon Telescope (EHT) Collaboration in 2019 \cite{eht,fbhs1,fbhs2,fbhs3,fbhs4,fbhs5,fbhs6}, which opens a new era in the fields of astrophysics and black hole physics. Nowadays, more and more researchers have devoted themselves to the study of black hole shadows. Black hole shadow is the dark region in the center of black hole image, which corresponds to the light rays fall into the event horizon of black hole. It plays a vital role in the study of black holes and verification of various gravity theories, because the fingerprints of the geometry around the black hole could be reflected in the shape and size of black hole shadow\cite{sha2,sha3}. The shadow of a Schwarzschild black hole is a perfect black disk. Kerr black hole shadow becomes a D-shaped silhouette gradually with the increase of spin parameter\cite{sha2,sha3}. The cusp silhouette of black hole shadows emerge in the spacetime of a Kerr black hole with Proca hair and a Konoplya-Zhidenko rotating non-Kerr black hole \cite{fpos2,sb10}. The eyebrow-like shadow and the self-similar fractal structures appear in the shadow for the disformal Kerr black hole in quadratic degenerate higher-order scalar-tensor theory \cite{lf}. In dS or AdS spacetimes, the size and shape of black hole shadow are related to the cosmological constant and the position of comoving observer \cite{xm31, xm32, xm33, xm34}. Many other black hole shadows with other parameters in various theories of gravity have been recently investigated in Refs. \cite{sw,swo,astro,chaotic,binary,sha18,my,sMN,swo7,mbw,mgw,sha4,sha5,sha6,sha7,sha8,sha9,sha10,sha11,sha111,sha12,sha13,sha14,sha141,sha15,sha16,sb1,sha17,sha19,sha191,sha192,sha193,sha194,shan1,shan1add,shan2add,shan3add,drk,rr,pe,lf2,Zeng2020vsj,Zeng2020dco}. These information imprinted in black hole shadow is believed widely to be captured in the future astronomical observations including the upgraded Event Horizon Telescope and BlackHoleCam\cite{bhc}.

In real astrophysical situations, an external magnetic field may exist around supermassive black holes, especially at the center of galaxies. It is believed that there are magnetic fields around M87 and Sgr A$^{*}$ black hole\cite{fbhs8, sgrac}. Therefore, it is much interesting to research the shadow of black hole immersed in an external magnetic field. In 1976, F. J. Ernst et al. obtain the solutions of Schwarzschild, Reissner-Nordstr$\ddot{o}$m and Kerr-Newman black holes embedded in Melvin universe \cite{binm8,binm9} where there exists a cylindrical symmetry magnetic field  aligned along the symmetry axis of black hole and it is described by the parameter $B$ \cite{melv, binm}, These static or axially symmetric solutions satisfy the Einstein-Maxwell equations with the magnetic field. Due to the existence of magnetic field, the geodesic equation in these solutions can not be variable-separated. Therefore, chaotic motion of test particle appears in Melvin-Schwarzschild black hole spacetime \cite{chinm}. Recently, H. C. D. Lima Junior et al investigated the shadow and chaotic lensing of Melvin-Schwarzschild black hole, and found the shadow could become a panoramic (equatorial) shadow when the magnetic field parameter $B$ is bigger than a critical value $B_{c}$\cite{slms}. It is natural to ask what new effects in black hole shadow arising from the spin parameter for a rotating black hole immersed in uniform  magnetic field. The main purpose of this paper is to probe the shadows of Kerr black hole immersed in magnetic field and to detect the effects of Melvin magnetic filed on the rotating black hole shadow.

The paper is organized as follows. In section II, we briefly review the spacetime of Melvin-Kerr black hole and study the unstable, stable light rings. In Section III, we present numerically the shadows of Melvin-Kerr black hole shadow with different magnetic field parameters $B$ and rotation parameter $a$, analyse the new features of Melvin-Kerr black hole shadow arising from the Melvin magnetic filed. Finally, we end the paper with a summary.

\section{The spacetime of Melvin-Kerr black hole and light rings}

In this section we consider a rotating black hole embedded in Melvin magnetic field, which is so called Melvin-Kerr black hole and can be expressed as \cite{binm9}
\begin{eqnarray}
\label{mkdg}
ds^{2}=|\Lambda|^{2}\Sigma [-\frac{\Delta}{\mathcal{A}}dt^{2}+\frac{dr^{2}}{\Delta}+d\theta^{2}]+\frac{\mathcal{A}}{\Sigma|\Lambda|^{2}}\sin^{2}\theta(d\phi-\omega dt)^{2},
\end{eqnarray}
where
\begin{eqnarray}
\label{mkdg1}
\Sigma &=& r^{2}+a^{2} \cos^{2}\theta,\;\;\;\;\;\;\;\;\;\;\; \mathcal{A}=(r^{2}+a^{2})^{2}-\Delta a^{2} \sin^{2}\theta, \nonumber\\ \;\;\;\;\;\;\
\Delta &=& r^{2}-2 Mr+a^{2}, \;\;\;\;\;\;\;\;\; \Lambda=1+\frac{1}{4}B^{2}\frac{\mathcal{A}}{\Sigma}\sin^{2}\theta-\frac{i}{2}B^{2}Ma \cos\theta (3-\cos^{2}\theta+\frac{a^{2}}{\Sigma}\sin^{4}\theta),
\end{eqnarray}
and
\begin{eqnarray}
\label{mkdg2}
\omega&=&\frac{a}{r^{2}+a^{2}}\Bigg\{(1-B^{4}M^{2}a^{2})-\Delta\Bigg[\frac{\Sigma}{\mathcal{A}}+\frac{B^{4}}{16}\Bigg(-8Mr\cos^{2}\theta(3-\cos^{2}\theta)-6Mr\sin^{4}\theta\nonumber\\
&+&\frac{2Ma^{2}\sin^{6}\theta}{\mathcal{A}}[r(r^{2}+a^{2})+2Ma^{2}]+\frac{4M^{2}a^{2}\cos^{2}\theta}{\mathcal{A}}[(r^{2}+a^{2})(3-\cos^{2}\theta)^{2}-4a^{2}\sin^{2}\theta]\Bigg)\Bigg]\Bigg\}.
\end{eqnarray}
The metric of Melvin-Kerr black hole is an axially symmetric solution of the Einstein-Maxwell equations with the magnetic field aligned along the symmetry axis of black hole\cite{binm8,binm9}. Here the parameters $M, a$ and $B$ denote the mass, the rotation parameter of black hole and the magnetic field parameter, respectively. For the magnetic field parameter $B = 0$, this metric (\ref{mkdg}) will reduce to Kerr solution. And for the rotation parameter $a=0$, the metric (\ref{mkdg}) will reduce to Melvin-Schwarzschild solution, which describes a Schwarzschild black hole immersed in Melvin magnetic field. The Hamiltonian $\mathcal{H}$ of a photon propagation along null geodesics in a Melvin-Kerr black hole spacetime (\ref{mkdg}) can be expressed as
\begin{eqnarray}
\label{lglr}
\mathcal{H}=\frac{1}{2}g^{\mu\nu}p_{\mu}p_{\nu}=0.
\end{eqnarray}
It is obvious that the coefficients of this metric are independent of the coordinates $t$ and $\phi$, which yields two constants of motion of the geodesics, i.e., the energy $E$ and the $z$-component of the angular momentum $L_{z}$,
\begin{eqnarray}
\label{EL}
E&=&-p_{t}=(|\Lambda|^{2}\Sigma \frac{\Delta}{\mathcal{A}}-\frac{\mathcal{A}}{|\Lambda|^{2}\Sigma}\omega^{2}\sin^{2}\theta)\dot{t}+\frac{\mathcal{A}}{|\Lambda|^{2}\Sigma}\omega\sin^{2}\theta\dot{\phi},\\ \
L_{z}&=&p_{\phi}=\frac{\mathcal{A}}{|\Lambda|^{2}\Sigma}\sin^{2}\theta\dot{\phi}-\frac{\mathcal{A}}{|\Lambda|^{2}\Sigma}\omega\sin^{2}\theta\dot{t}.
\end{eqnarray}

The Hamiltonian $\mathcal{H}$ can be rewritten as
\begin{eqnarray}
\label{hmd}
\mathcal{H}=g^{rr}p_{r}^{2}+g^{\theta\theta}p_{\theta}^{2}+V_{eff},
\end{eqnarray}
where the effective potential $V_{eff}$ is defined as
\begin{eqnarray}
\label{veff}
V_{eff}=g^{\phi\phi}E^{2}(\eta^{2}+2\frac{g_{t\phi}}{g_{tt}}\eta+\frac{g_{\phi\phi}}{g_{tt}}),
\end{eqnarray}
and $\eta=L/E$ is the impact parameter.

Now, let us study the circular photon orbits in the equatorial plane also known as light rings. From the Hamiltonian $\mathcal{H}$ (\ref{hmd}), we can find that the light rings must satisfy both $\theta=\pi/2$ and
\begin{eqnarray}
\label{lr}
V_{eff}=0,\;\;\;\;\;\;\;\;\;\;\;\;\;\;\;\;\;\;\;\;\;\frac{\partial V_{eff}}{\partial r}=0.
\end{eqnarray}
Moreover, when $\partial^{2}V_{eff}/\partial r^{2}<0$, the light ring is an unstable circular photon orbit, which determines the black hole shadow boundary in the equatorial plane. When $\partial^{2}V_{eff}/\partial r^{2}>0$, the light ring is stable. Generally, in Schwarzschild or Kerr black hole spacetime there is no stable light ring. The interesting thing for Melvin-Kerr black hole spactime is there could exist stable light ring. For $V_{eff}=0$ one can obtain
\begin{eqnarray}
\label{et12}
\eta_{1}=-\frac{g_{t\phi}}{g_{tt}}-\frac{\sqrt{g_{t\phi}^{2}-g_{tt}g_{\phi\phi}}}{g_{tt}},\;\;\;\;\;\;\;\;\;\;\;\;\;\;\;\;\;\;\;\;\; \eta_{2}=-\frac{g_{t\phi}}{g_{tt}}+\frac{\sqrt{g_{t\phi}^{2}-g_{tt}g_{\phi\phi}}}{g_{tt}},
\end{eqnarray}
\begin{figure}
\includegraphics[width=16cm ]{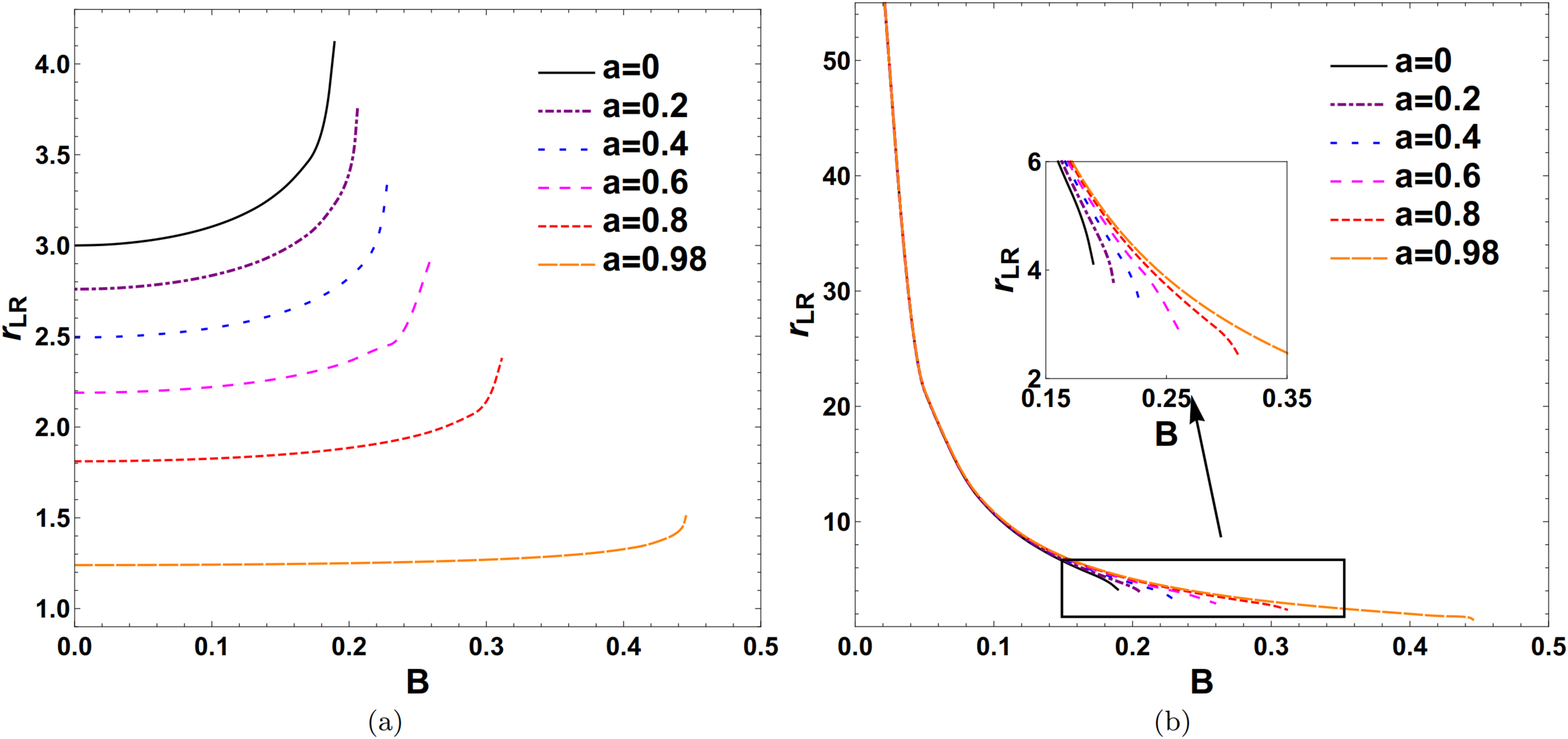}
\caption{(a) Varieties of the prograde unstable light rings radiuses $r_{LR}$ with the rotation parameter $a$ and the magnetic field parameter $B$. (b) Varieties of the prograde stable light rings radiuses $r_{LR}$ with the magnetic field parameter $B$ and the rotation parameter $a$. Here we set M=1.}
\label{retro}
\end{figure}
\begin{figure}
\includegraphics[width=16cm ]{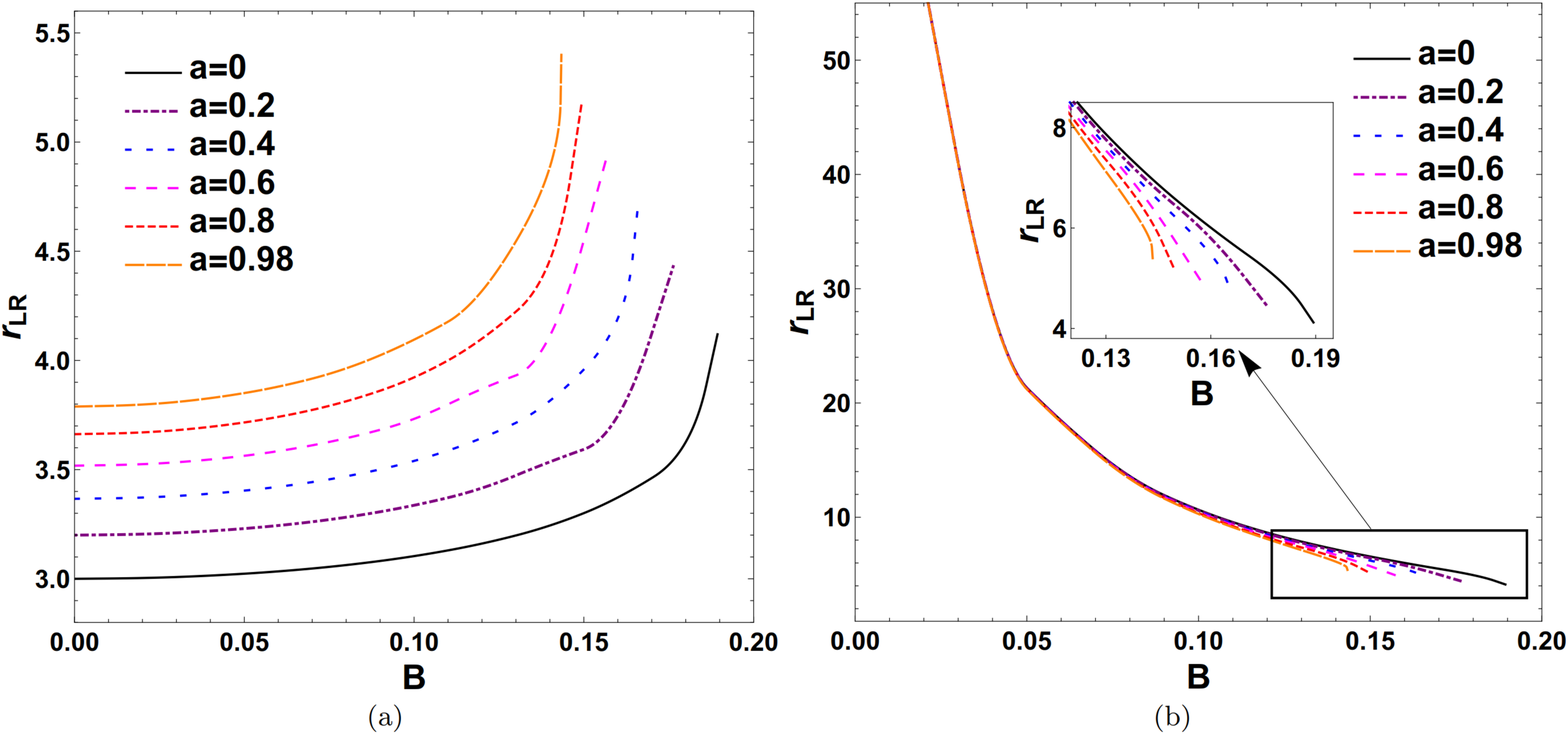}
\caption{(a) Varieties of the retrograde unstable light rings radiuses $r_{LR}$ with the rotation parameter $a$ and the magnetic field parameter $B$. (b) Varieties of the retrograde stable light rings radiuses $r_{LR}$ with the magnetic field parameter $B$ and the rotation parameter $a$. Here we set M=1.}
\label{pro}
\end{figure}
\begin{figure}
\includegraphics[width=8cm ]{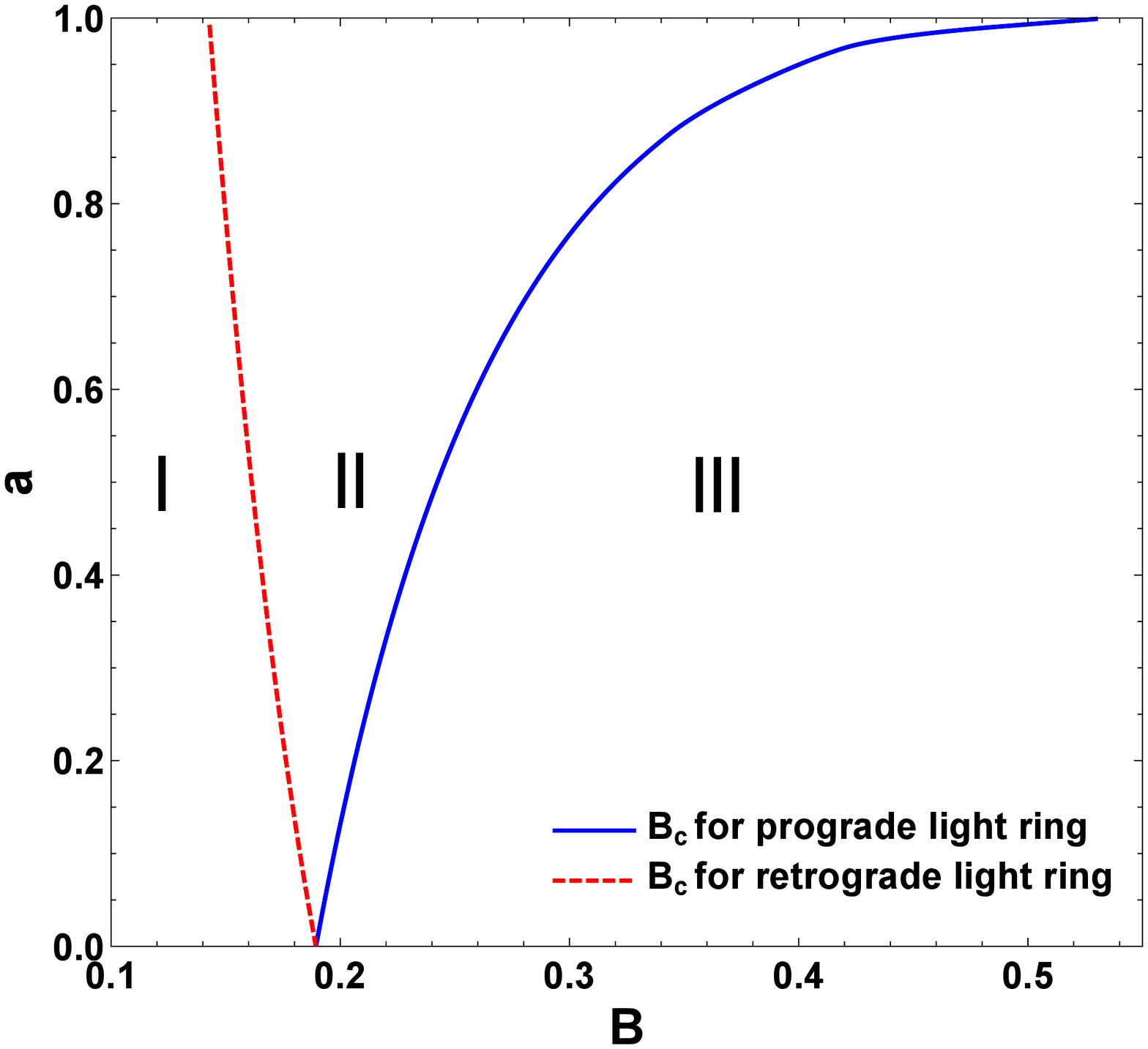}
\caption{The critical value $B_{c}$ for prograde and retrograde light rings with different rotation parameter $a$ in ($B$, $a$) plane. The regions $I$, $II$ and $III$ are separated by curves of $B_{c}$ for prograde and retrograde light rings in the panel. In region $I$, the prograde light ring and the retrograde light ring both exist in Melvin-Kerr black hole spacetime. In region $II$, only prograde light ring exists. In region $III$, neither the prograde light ring nor the retrograde light ring exist. Here we set M = 1.}
\label{bc}
\end{figure}
which represent the impact parameter $\eta$ of the prograde light ring and the retrograde light ring respectively when seen from an observer on the equatorial plane. In Fig.\ref{retro}-\ref{pro}, we show the varieties of the prograde and retrograde unstable (stable) light rings radiuses $r_{LR}$ with the magnetic field parameter $B$ and the rotation parameter $a$. One can find both the prograde and retrograde unstable light rings radiuses $r_{LR}$ increase with the magnetic field parameter $B$, but it is the opposite for stable light rings. When $B$ approaches $0$ (Kerr spacetime), the stable light ring radius $r_{LR}$ approaches infinity. For the prograde light rings, the radius $r_{LR}$ of unstable light ring is smaller for bigger rotation parameter $a$, and the stable light ring radius $r_{LR}$ is bigger for bigger $a$. However, for the retrograde light rings, the unstable light ring radius $r_{LR}$ is bigger for bigger $a$, the stable light ring radius $r_{LR}$ is smaller for bigger $a$. For each rotation parameter $a$, there is a critical value of magnetic field parameter $B_{c}$, that is the value of $B$ in the rightmost point of the variety curve of light ring radius $r_{LR}$. The unstable and stable light rings would disappear after the same $B_{c}$. However, the critical value of magnetic field parameter $B_{c}$ for prograde light ring is generally bigger than that for retrograde light ring. Fig.\ref{bc} exhibits the critical value $B_{c}$ for prograde and retrograde light rings with different rotation parameter $a$ in ($B$, $a$) plane. The regions $I$, $II$ and $III$ are separated by curves of $B_{c}$ for prograde and retrograde light rings in the panel. In region $I$, the prograde light ring and the retrograde light ring both exist in Melvin-Kerr black hole spacetime. In region $II$, only prograde light ring exists. In region $III$, neither the prograde light ring nor the retrograde light ring exist.

\section{The shadows of of Melvin-Kerr black hole}

Now, we study the shadows of Melvin-Kerr black hole through the backward ray-tracing technique \cite{sw,swo,astro,chaotic,binary,sha18,my,sMN,swo7,mbw}, using the fifth-order Runge-Kutta method. We assume that the static observer is locally at ($r_{obs}, \theta_{obs}$) in zero-angular-moment-observers reference frame (ZAMOs) \cite{Bardeen}, and then the observer basis $\{e_{\hat{t}},e_{\hat{r}},e_{\hat{\theta}},e_{\hat{\phi}}\}$ can be expanded as a form in the coordinate basis $\{\partial_t,\partial_r,\partial_{\theta},\partial_{\phi} \}$ \cite{sw,swo,astro,chaotic,binary,sha18,my,sMN,swo7,mbw}
\begin{eqnarray}
\label{zbbh}
e_{\hat{\mu}}=e^{\nu}_{\hat{\mu}} \partial_{\nu},
\end{eqnarray}
where the transform matrix $e^{\nu}_{\hat{\mu}}$ obeys to $g_{\mu\nu}e^{\mu}_{\hat{\alpha}}e^{\nu}_{\hat{\beta}}
=\eta_{\hat{\alpha}\hat{\beta}}$, and $\eta_{\hat{\alpha}\hat{\beta}}$ is the metric of Minkowski spactime.
For the Melvin-Kerr spacetime (\ref{mkdg}), it is convenient to choice a decomposition\cite{sw,swo,astro,chaotic,binary,sha18,my,sMN,swo7,mbw,mgw}
\begin{eqnarray}
\label{zbbh1}
e^{\nu}_{\hat{\mu}}=\left(\begin{array}{cccc}
\zeta&0&0&\gamma\\
0&A^r&0&0\\
0&0&A^{\theta}&0\\
0&0&0&A^{\phi}
\end{array}\right),
\end{eqnarray}
where $\zeta$, $\gamma$, $A^r$, $A^{\theta}$, and $A^{\phi}$ are real coefficients.
From the Minkowski normalization
\begin{eqnarray}
e_{\hat{\mu}}e^{\hat{\nu}}=\delta_{\hat{\mu}}^{\hat{\nu}},
\end{eqnarray}
one can obtain
\begin{eqnarray}
\label{xs}
&&A^r=\frac{1}{\sqrt{g_{rr}}},\;\;\;\;\;\;\;\;\;\;\;\;\;\;\;\;
A^{\theta}=\frac{1}{\sqrt{g_{\theta\theta}}},\;\;\;\;\;\;\;\;\;\;\;\;\;\;\;
A^{\phi}=\frac{1}{\sqrt{g_{\phi\phi}}},\nonumber\\
&&\zeta=\sqrt{\frac{g_{\phi \phi}}{g_{t\phi}^{2}-g_{tt}g_{\phi \phi}}},\;\;\;\;\;\;\;\;\;\;\;\;\;\;\;\;\;\;\;\; \gamma=-\frac{g_{t\phi}}{g_{\phi\phi}}\sqrt{\frac{g_{\phi \phi}}{g_{t\phi}^{2}-g_{tt}g_{\phi \phi}}}.
\end{eqnarray}
\begin{figure}
\includegraphics[width=16cm ]{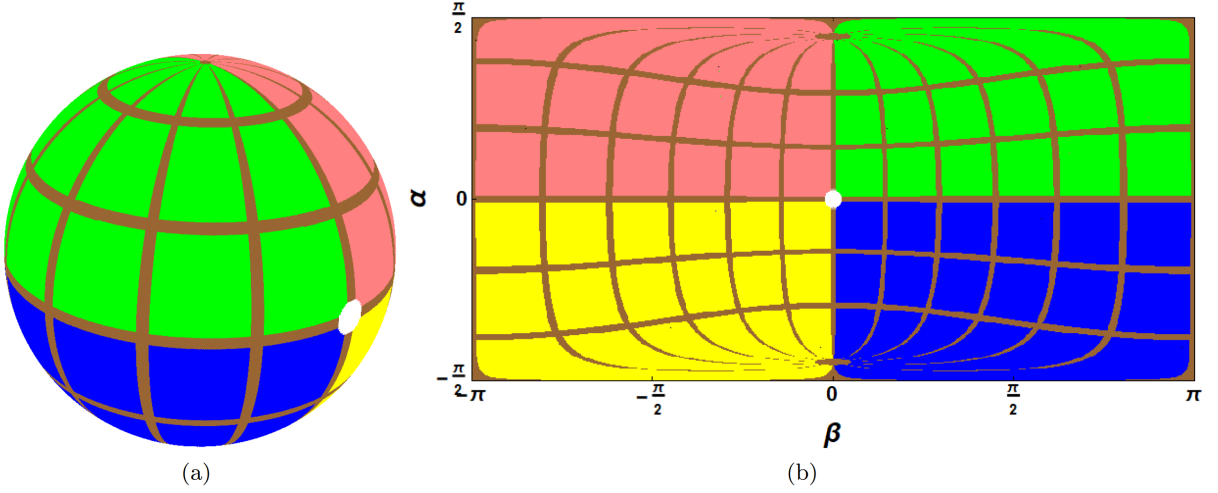}
\caption{(a) The sphere light source marked by four different colored quadrants, the brown grids of longitude and latitude, and the white reference spot lies at the intersection of the four colored quadrants. (b) The panoramic image of the sphere light source for observer at $r_{obs}=8$ and $r_{sphere}=50$ in Minkowski spacetime.}
\label{bj}
\end{figure}
Therefore, one can get the locally measured four-momentum $p^{\hat{\mu}}$ of a photon by the projection of its four-momentum $p^{\mu}$  onto $e_{\hat{\mu}}$,
\begin{eqnarray}
\label{dl}
p^{\hat{t}}=-p_{\hat{t}}=-e^{\nu}_{\hat{t}} p_{\nu},\;\;\;\;\;\;\;\;\;
\;\;\;\;\;\;\;\;\;\;\;p^{\hat{i}}=p_{\hat{i}}=e^{\nu}_{\hat{i}} p_{\nu}.
\end{eqnarray}
With the help of Eq.(\ref{xs}), the locally measured four-momentum $p^{\hat{\mu}}$ can be further written as
\begin{eqnarray}
\label{kmbh}
p^{\hat{t}}&=&\zeta E-\gamma L,\;\;\;\;\;\;\;\;\;\;\;\;\;\;\;\;\;\;\;\;p^{\hat{r}}=\frac{1}{\sqrt{g_{rr}}}p_{r} ,\nonumber\\
p^{\hat{\theta}}&=&\frac{1}{\sqrt{g_{\theta\theta}}}p_{\theta},
\;\;\;\;\;\;\;\;\;\;\;\;\;\;\;\;\;\;\;\;\;\;
p^{\hat{\phi}}=\frac{1}{\sqrt{g_{\phi\phi}}}L.
\end{eqnarray}
\begin{figure}
\includegraphics[width=16cm ]{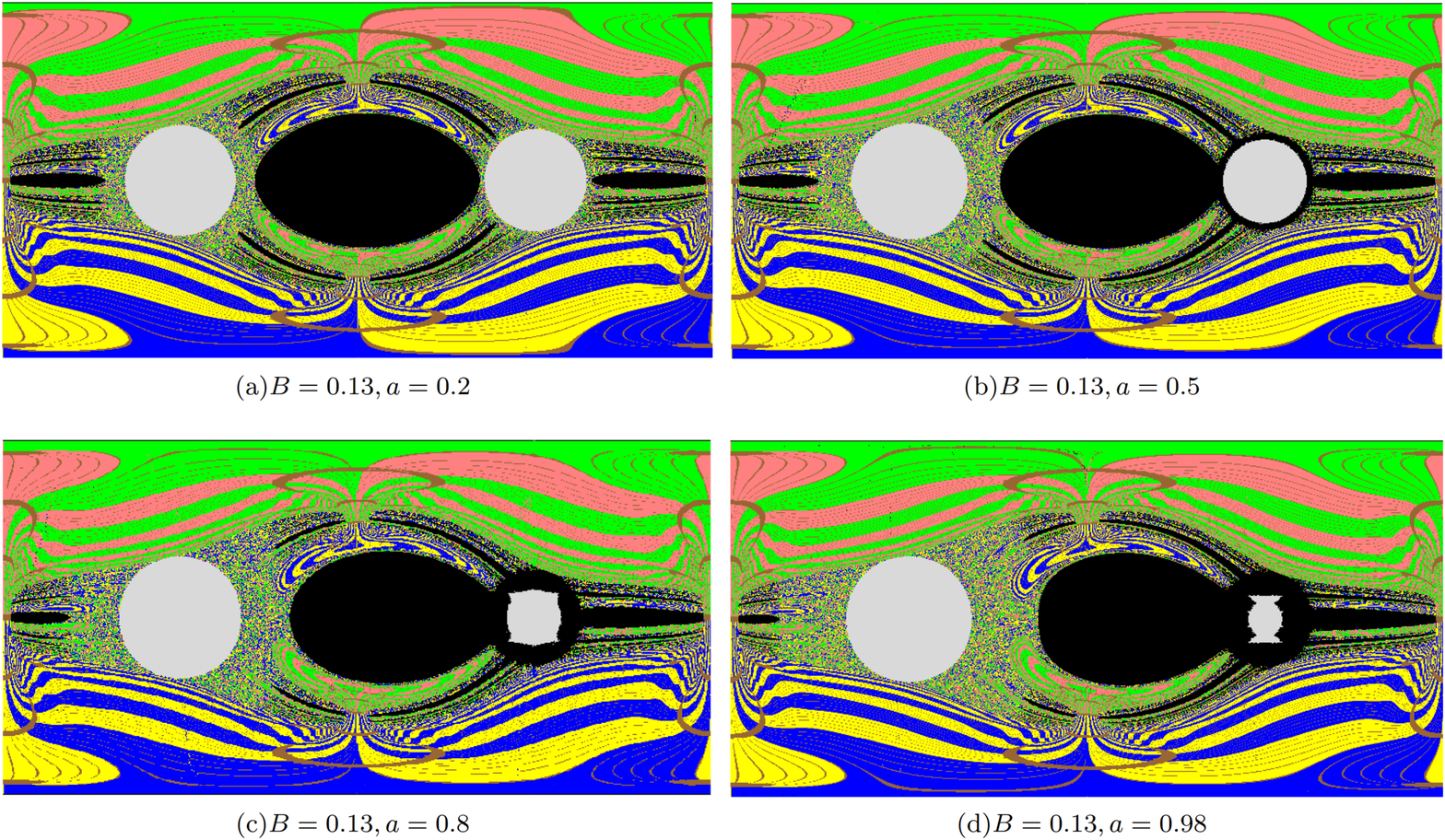}
\caption{The shadows of Melvin-Kerr black hole with different rotation parameter $a$ and the magnetic field parameter $B=0.13$. Here we set $M=1$, $r_{sphere}=50$ and the observer at $r_{obs}=8$ with the inclination angle $\theta_{obs}=\pi/2$.}
\label{13}
\end{figure}
\begin{figure}
\includegraphics[width=16cm ]{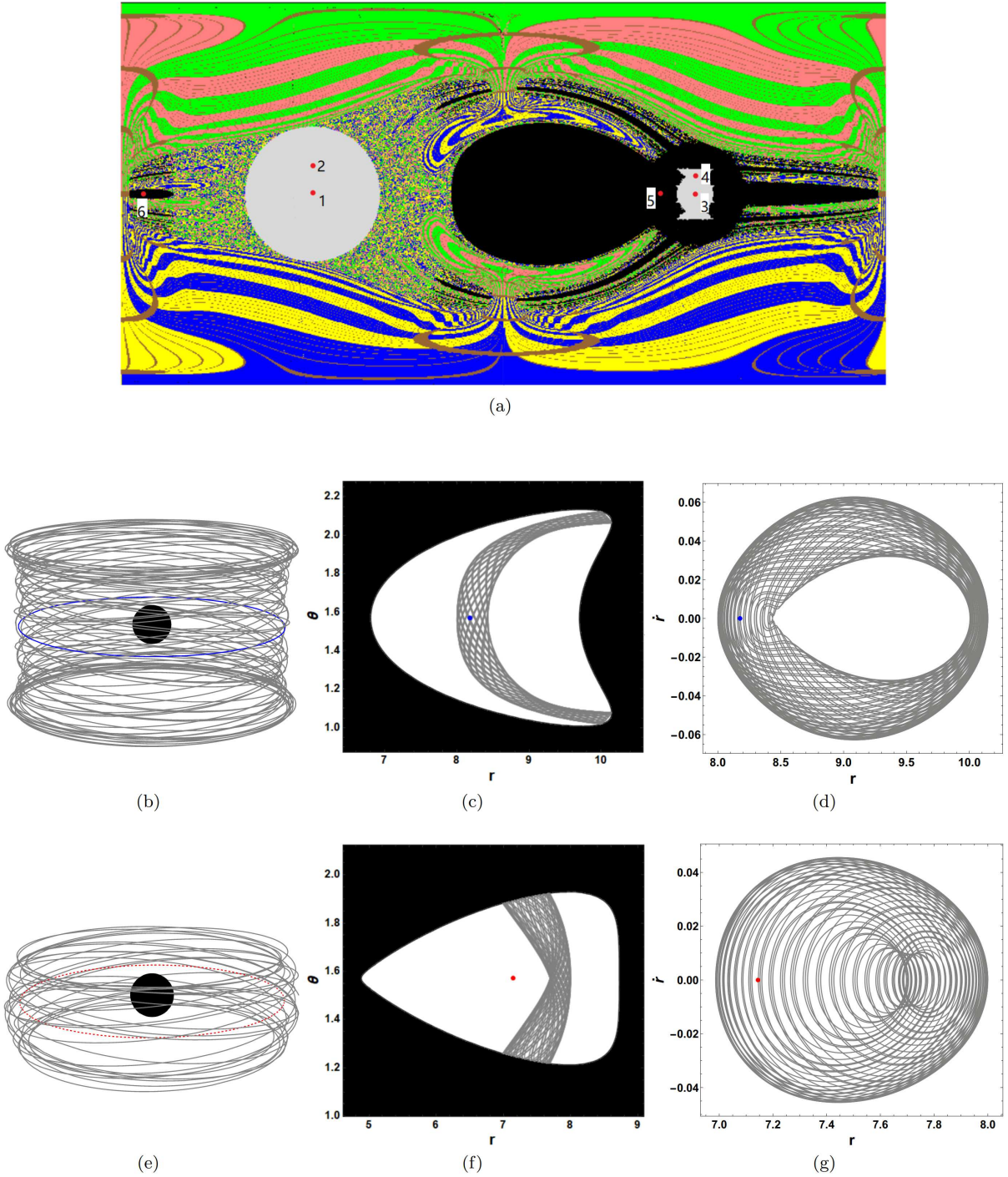}
\caption{Top: The shadow of Melvin-Kerr black hole with the rotation parameter $a=0.98$ and the magnetic field parameter $B=0.13$, which is same as Fig.\ref{13}(d). Middle: The prograde stable photon orbit corresponding to the point $2$ in figure $a$ with the impact parameter $\eta=5.48$, The blue circle or dot represents the prograde stable light ring. Bottom: The retrograde stable photon orbit corresponding to the point $4$ in figure $a$ with the impact parameter $\eta=-5.81$, The red dotted circle or dot represents the retrograde stable light ring.}
\label{dt}
\end{figure}
\begin{figure}
\includegraphics[width=12cm ]{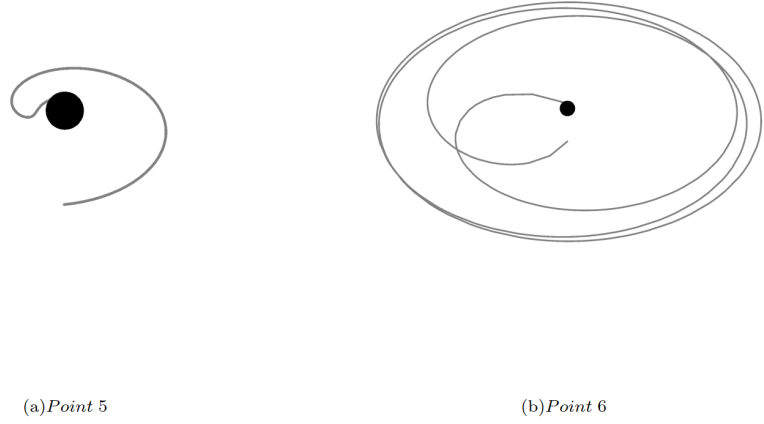}
\caption{The photon trajectories corresponding to points $5, 6$ in Melvin-Kerr black hole shadow in Fig.\ref{dt}(a). The dark spheres represent Melvin-Kerr black hole.}
\label{56}
\end{figure}
Repeating the similar operations in Refs.\cite{sw,swo,astro,chaotic,binary,sha18,my,sMN,swo7,mbw},
the angular coordinates ($\alpha$, $\beta$) in the observer's local sky define the direction of the associated light ray can be expressed as \cite{sw,swo,astro,chaotic,binary,sha18,my,sMN,swo7,mbw,mgw}
\begin{eqnarray}
\label{xd1}
\beta=-\arctan\frac{p^{\hat{\phi}}}{p^{\hat{r}}}|_{(r_{obs},\theta_{obs})}, \;\;\;\;\;\;\;\;\;\;\;\;\;\;\;\;\;\;
\alpha=\arcsin\frac{p^{\hat{\theta}}}{p^{\hat{t}}}|_{(r_{obs},\theta_{obs})}.
\end{eqnarray}

We set a sphere light source marked by four different colored quadrants with radius $r_{sphere}$, the brown grids as longitude and latitude, and the white reference spot lies at the intersection of the four colored quadrants, which is same as the celestial sphere in Ref.\cite{my}, shown in Fig.\ref{bj}(a). Melvin-Kerr black hole is placed at the center of the light sphere. The observer is placed along the axis joining the sphere center and the other intersection of the four colored quadrants with $r_{obs}<r_{sphere}$. The panoramic image of the sphere light source for observer at $r_{obs}=8$ and $r_{sphere}=50$ in Minkowski spacetime is shown in Fig.\ref{bj}(b). The angular coordinates $-\pi\leq\beta\leq\pi$, $-\pi/2\leq\alpha\leq\pi/2$.
\begin{figure}
\includegraphics[width=16cm ]{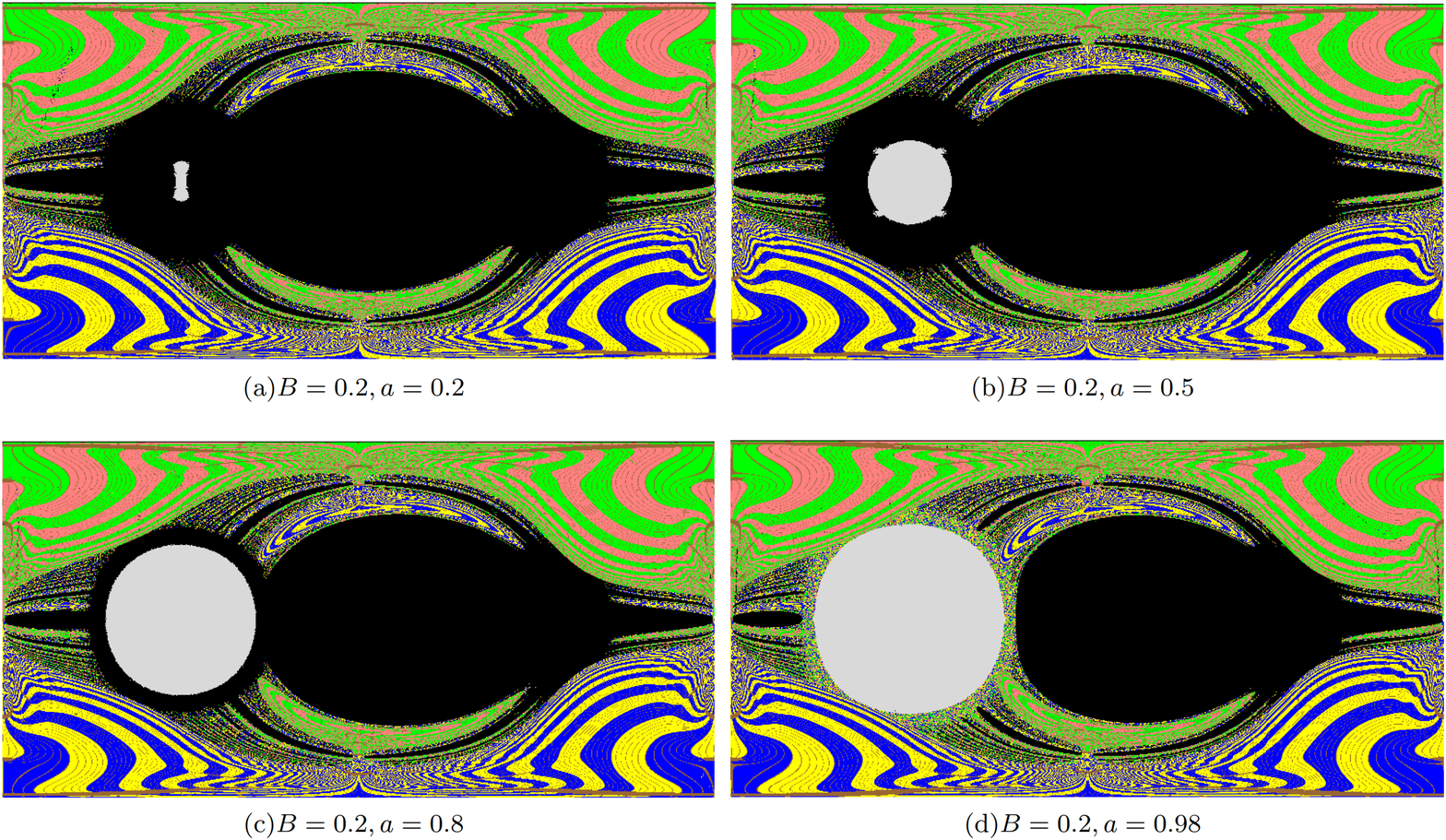}
\caption{The shadows of Melvin-Kerr black hole with different rotation parameter $a$ and the magnetic field parameter $B=0.2$. Here we set $M=1$, $r_{sphere}=50$ and the observer at $r_{obs}=4.8$ with the inclination angle $\theta_{obs}=\pi/2$.}
\label{2}
\end{figure}
\begin{figure}
\includegraphics[width=13cm ]{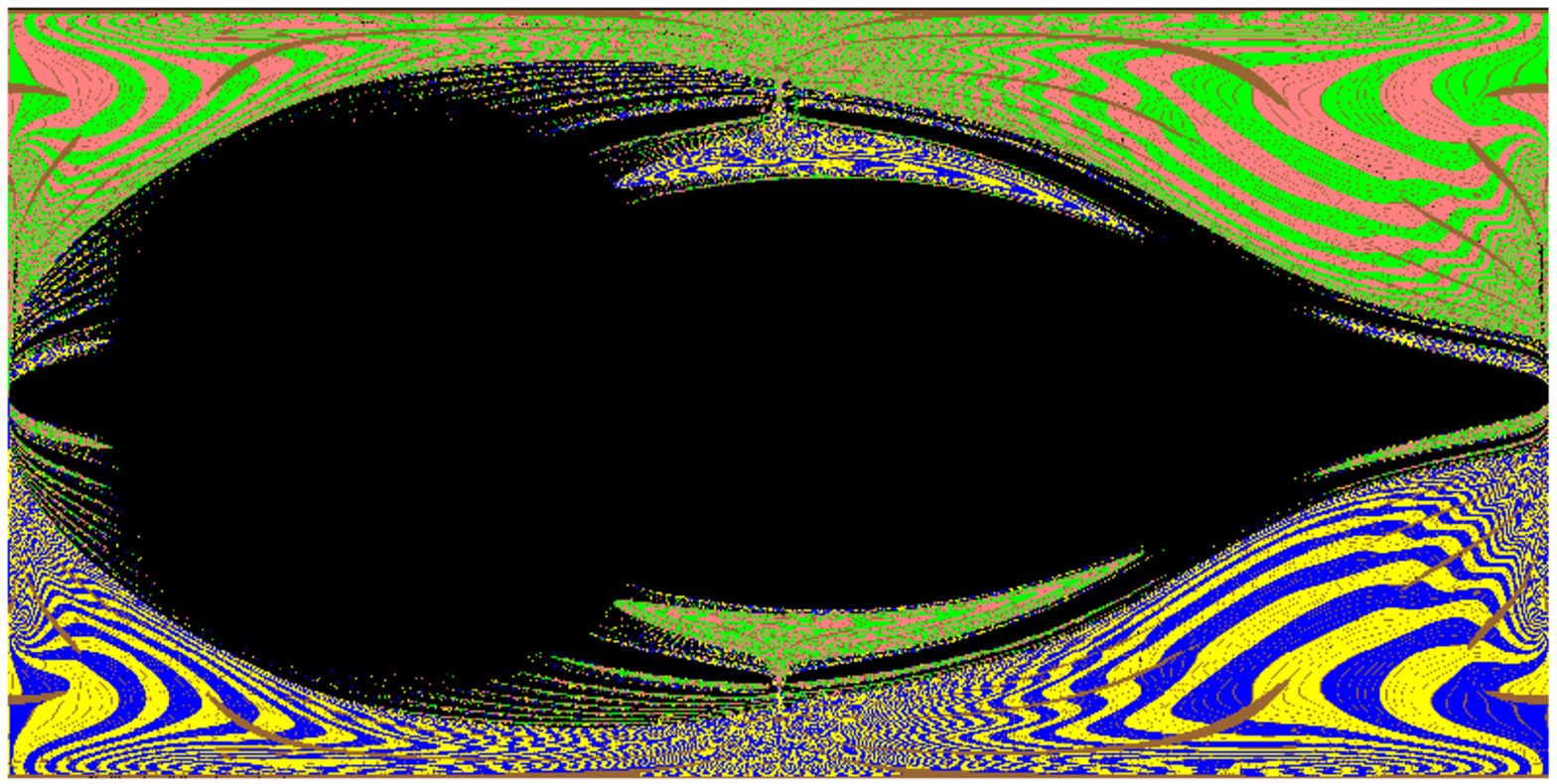}
\caption{The shadows of Melvin-Kerr black hole with the rotation parameter $a=0.8$, the magnetic field parameter $B=0.35$. Here we set $M=1$, $r_{sphere}=50$ and the observer at $r_{obs}=4.8$ with the inclination angle $\theta_{obs}=\pi/2$.}
\label{35}
\end{figure}

In Fig.\ref{13}, we present the shadows of Melvin-Kerr black hole with different rotation parameter $a$, the magnetic field parameter $B=0.13$, and we set $M=1$, $r_{sphere}=50$ and the observer at $r_{obs}=8$ with the inclination angle $\theta_{obs}=\pi/2$. The value of $B$ and $a$ we set in this figure allows both the prograde and retrograde unstable (stable) light rings exist. This figure also is a panoramic image that the observer observed ($-\pi\leq\beta\leq\pi$, $-\pi/2\leq\alpha\leq\pi/2$). One can find the multiple disconnected shadows of Melvin-Kerr black hole including a middle oblate shadow and much striped shadows, which possesses self-similar fractal structures originating from the chaotic lensing. With the increasing of the rotation parameter $a$, the middle main shadow gradually shifts to the right side, which is similar to Kerr black hole shadow. The new features in Melvin-Kerr black hole shadow is the two gray regions on both sides of the middle main shadow. They emerge as a result of there are prograde and retrograde stable photon orbits around the stable light rings, some of them are outside the equatorial plane. The points $1, 3$ in Fig.\ref{dt}(a) (same as Fig.\ref{13}(d)) correspond to the prograde and retrograde stable light rings respectively, the trajectories of them are shown as the blue and red dotted circles in Fig.\ref{dt}(b) and (e) respectively. The points $2, 4$ in Fig.\ref{dt}(a) correspond to the prograde and retrograde stable photon orbits with the impact parameter $\eta=5.48, -5.81$ respectively, the trajectories of them are shown as the gray curves in Fig.\ref{dt}(b) and (e). We found the photons in stable orbits are always moving around Melvin-Kerr black hole, the dark spheres in Fig.\ref{dt}(b) and (e). Fig.\ref{dt}(c) and (f) exhibit the effective potential $V_{eff}=0$ contour plot with $\eta=5.48, -5.81$ respectively, the dark regions ($V_{eff}>0$) are the forbidden regions for photons. The gray curves in Fig.\ref{dt}(c) and (f) represent the stable photon orbits in ($r, \theta$) plane for points $2, 4$. One can find the stable photon orbits are trapped in the region $V_{eff}\leq0$, they can't even enter the black hole. So the gray region doesn't belong to the black hole shadow, but if there is no light sources in the stable photon orbits region the observer also will see dark shadow in the gray region. The blue and red dots in Fig.\ref{dt}(c), (d), (f) and (g) represent the prograde and retrograde stable light rings respectively. Fig.\ref{dt}(d) and (g) exhibit the prograde and retrograde stable photon orbits in ($r, \dot{r}$) plane. What's more, as the rotation parameter $a$ increases, the gray region becomes smaller, which means the observer will see less stable photon orbits for bigger $a$ in this case. Fig.\ref{56} exhibits the photon trajectories for points $5, 6$ in Melvin-Kerr black hole shadow in Fig.\ref{dt}(a). In Fig.\ref{dt}(a) the point $5$ locates on the inner side of the gray region (stable photon orbit region), and we find the photon for point $5$ cannot orbit the black hole one circle in Fig.\ref{56}(a). However, we can find the photon for point $6$ locating on outer side of the gray region moves a few orbits before it enters Melvin-Kerr black hole in Fig.\ref{56}(b).

In Fig.\ref{2}, we present the shadows of Melvin-Kerr black hole with different rotation parameter $a$, the magnetic field parameter $B=0.2$, and we set $M=1$, $r_{sphere}=50$ and the observer at $r_{obs}=4.8$ with the inclination angle $\theta_{obs}=\pi/2$. The value of $B$ and $a$ we set in this figure only allows the prograde unstable (stable) light ring exists. One can find the gray region only emerges in the life of the main shadow, which corresponds to the prograde stable photon orbits. As the rotation parameter $a$ increases, the gray region becomes bigger, which means the observer could see more stable photon orbits for bigger $a$ in this case. The absence of the retrograde unstable (stable) light rings makes the Melvin-Kerr black hole shadow an half-panoramic (equatorial) shadow which spans the entire right half of the field of observer view ($0\leq\beta\leq\pi$). It is consistent with Ref.\cite{slms}, the absence of equatorial light rings yields a panoramic shadow of Melvin-Schwarzschild black hole. It is because that the unstable light ring determines the black hole shadow boundary in the equatorial plane.

In Fig.\ref{35}, we present the shadows of Melvin-Kerr black hole with the rotation parameter $a=0.8$, the magnetic field parameter $B=0.35$, and we set $M=1$, $r_{sphere}=50$ and the observer at $r_{obs}=4.8$ with the inclination angle $\theta_{obs}=\pi/2$. The value of $B$ and $a$ we set in this figure don't allows neither the prograde nor retroprade unstable (stable) light ring exists. So we can't see the gray region representing stable photon orbits in this figure. Due to the absence of the light rings, Melvin-Kerr black hole shadow becomes a panoramic (equatorial) shadow which spans the entire field of observer view ($-\pi\leq\beta\leq\pi$). In addition, the rotation of Melvin-Kerr black hole makes the left side of the shadow larger than the right side.

\section{summary}

We have studied the spacetime of a Kerr black hole immersed in Melvin magnetic field, and found not only unstable light rings could exist, but also stable light rings could exist. Both the prograde and retrograde unstable light rings radiuses increase with the magnetic field parameter $B$, but it is the opposite for stable light rings. For the prograde light ring, the radius of unstable light ring is smaller for bigger rotation parameter $a$, and the stable light ring radius is bigger for bigger $a$. However, for the retrograde light ring, the unstable light ring radius is bigger for bigger $a$, and the stable light ring radius is smaller for bigger $a$. The existence of unstable, stable light rings depend on both the rotation parameter $a$ and the magnetic field parameter $B$. For a certain $a$, there are both the prograde and retroprade unstable (stable) light rings when $B$ is less than a critical value of magnetic field parameter $B_{c}$ of retrograde light ring. In this case, the shadows of Melvin-Kerr black hole have two gray regions on both sides of the middle main shadow, which correspond to the prograde and retrograde stable photon orbits. The photons in stable orbits are always moving around Melvin-Kerr black hole, they can't enter the black hole or escape to infinity. As $B$ continues to increase, there is only the prograde unstable (stable) light ring. In this case, the gray region only emerges in the life of the main shadow, which corresponds to the prograde stable photon orbits. The absence of the retrograde unstable (stable) light rings makes the Melvin-Kerr black hole shadow an half-panoramic (equatorial) shadow which spans the entire right half of the field of observer view ($0\leq\beta\leq\pi$). When $B$ is bigger than $B_{C}$ of prograde light ring, neither prograde nor retroprade unstable (stable) light rings exist. In this case, the shadow of Melvin-Kerr black hole has no gray region for stable photon orbits, and becomes a panoramic (equatorial) shadow which spans the entire field of observer view ($-\pi\leq\beta\leq\pi$). The self-similar fractal structures are also found in the shadows of Melvin-Kerr black hole, which are caused by the chaotic motion of photon. Our results show that the Melvin magnetic field yields novel effects on Kerr black hole shadow.

\section{\bf Acknowledgments}

This work was supported by the Shandong Provincial Natural Science Foundation of China under Grant No. ZR2020QA080, and was partially supported by the National Natural Science Foundation of China under Grant No. 11875026, 11475061, 11875025, and 12035005.

\end{document}